\RequirePackage{fix-cm}
\documentclass[]{pasj02} 
\usepackage{natbib} 
\usepackage{siunitx}
\usepackage{url} 
\jyear{2025}
\Received{}
\Accepted{}

\begin{document} 

\title{High count rate effects in event processing for XRISM/Resolve x-ray microcalorimeter: II. Energy scale and resolution in orbit}

\author{
Misaki \textsc{Mizumoto}\altaffilmark{1}\altemailmark\orcid{0000-0003-2161-0361}\email{mizumoto-m@fukuoka-edu.ac.jp},
 Yoshiaki \textsc{Kanemaru}\altaffilmark{2}\orcid{0000-0002-4541-1044}, 
 Shinya \textsc{Yamada}\altaffilmark{3}\orcid{0000-0003-4808-893X}, 
Caroline~A.~\textsc{Kilbourne}\altaffilmark{4}\orcid{0000-0001-9464-4103}, 
Megan~E.~\textsc{Eckart}\altaffilmark{5}\orcid{0000-0003-3894-5889},
Edmund \textsc{Hodges-Kluck}\altaffilmark{4}\orcid{0000-0002-2397-206X},
Yoshitaka \textsc{Ishisaki}\altaffilmark{6}\orcid{0000-0003-0163-7217},
Frederick S.~\textsc{Porter}\altaffilmark{4}\orcid{0000-0002-6374-1119},
Katja~\textsc{Pottschmidt}\altaffilmark{7,8}\orcid{0000-0002-4656-6881},
and Tsubasa \textsc{Tamba}\altaffilmark{2}\orcid{0000-0001-7631-4362} 
}
\altaffiltext{1}{Science Research Education Unit, University of Teacher Education Fukuoka, 1-1 Akama-bunkyo-machi, Munakata, Fukuoka 811-4192, Japan}
\altaffiltext{2}{Institute of Space and Astronautical Science (ISAS), Japan Aerospace Exploration Agency (JAXA), 3-1-1 Yoshinodai, Chuo-ku, Sagamihara, Kanagawa 252-5210, Japan}
\altaffiltext{3}{Department of Physics, Rikkyo University, 3-34-1 Nishi Ikebukuro, Toshima-ku, Tokyo 171-8501, Japan}
\altaffiltext{4}{NASA Goddard Space Flight Center, Greenbelt, MD 20771, USA}
\altaffiltext{5}{Lawrence Livermore National Laboratory, 7000 East Avenue, Livermore CA 94550, USA}
\altaffiltext{6}{Faculty of Science, Tokyo Metropolitan University,	1-1 Minami-Osawa, Hachioji, Tokyo 192-0397, Japan}
\altaffiltext{7}{University of Maryland Baltimore County, 1000 Hilltop Circle, Baltimore, MD 21250, USA}
\altaffiltext{8}{CRESST \& Astroparticle Physics Laboratory, NASA Goddard Space Flight Center, 8800 Greenbelt Road, Greenbelt, MD 20771, USA}



\maketitle

\begin{abstract}
The Resolve instrument on the X-ray Imaging and Spectroscopy Mission (XRISM) uses a 36-pixel microcalorimeter designed to deliver high-resolution, non-dispersive X-ray spectroscopy. {Although it is optimized for extended sources with low count rates,} Resolve observations of bright point sources are still able to provide unique insights into the physics of these objects, as long as high {count} rate effects are addressed in the analysis. 
These effects include {the loss of exposure time for each pixel}, change on the energy scale, and change on the energy resolution. To investigate these effects under realistic observational conditions, we observed the bright X-ray source, the Crab Nebula, with XRISM at several offset positions with respect to the Resolve field of view and with continuous illumination from {$^{55}$Fe sources} on the filter wheel. 
{For the spectral analysis, we excluded data where exposure time loss was too significant to ensure reliable spectral statistics.}
The energy scale at 6 keV shows a slight negative shift in the high-count-rate regime.
The energy resolution at 6 keV worsens as the count rate in electrically neighboring pixels increases, but can be restored by applying a nearest-neighbor coincidence cut (``cross-talk cut''). We examined how these effects influence the observation of bright point sources, using GX 13+1 as a test case, and identified an eV-scale energy offset at 6 keV between the inner (brighter) and outer (fainter) pixels. Users who seek to analyze velocity structures on the order of tens of km~s$^{-1}$ should account for such high count rate effects. These findings will aid in the interpretation of Resolve data from bright sources and provide valuable considerations for designing and planning for future microcalorimeter missions.
\end{abstract}


\section{Introduction}

The X-ray Imaging and Spectroscopy Mission (XRISM: \citealt{XRISM}), launched on 2023 September 6 in UT, aims to deliver groundbreaking advancements in X-ray astronomy. At the heart of XRISM lies the Resolve instrument \citep{ishisaki2022, Sato2023,kelley2024}, a 36-pixel microcalorimeter designed for high-resolution non-dispersive spectroscopy with an energy resolution of approximately 5~eV in the 0.3--12~keV range. An X-ray microcalorimeter is a device that converts the energy of incoming X-ray photons into heat, sensing the resulting temperature change in a sub-Kelvin-cooled thermometer \citep{McCammon1984}. Each pixel is individually coupled to the heat sink, and the signal time constant is determined by the thermal properties of that link. 
These features allow unprecedented energy resolution over a broad range without dispersion, making microcalorimeters indispensable for modern X-ray astronomy by enabling precise measurements of astrophysical phenomena.

{Some} astronomical sources are so bright ($\gtrsim0.2$~Crab) that the high rate of X-rays incident on the Resolve array poses a special challenge to the data analysis and scientific inference. 
XRISM investigations of these sources, such as the Crab Nebula or transient black hole binaries, are of great scientific interest, so it is important to understand the effect of high count rates on the performance of Resolve. 
Several effects can occur at high count rates: event loss caused when the CPU cannot fully process X-ray events as fast as they are detected, energy-scale shifts due to local heating by the X-ray flux, and distortion of event pulse shapes (used to infer the X-ray photon energy) by untriggered electrical cross-talk between electrically neighboring pixels. 
The onboard CPU (Pulse Shape Processor; PSP; \citealt{Ishisaki2018}), which was originally designed for the Hitomi \citep{takahashi2016}/Soft X-ray Spectrometer (SXS; \citealt{Kelley2016}) and was not changed for XRISM, 
is responsible for event processing in orbit.
The impact of each effect has been studied using ground-based testing \citep{mizumoto2025}. Here we address the in-orbit performance when observing the Crab Nebula.

In this study, we focus on energy-scale shifts and cross-talk effects, examining their impact on spectral distortion using in-orbit data. 
The Crab Nebula, a well-known bright X-ray source, was observed using Resolve while its Filter Wheel (FW) was rotated to an open position that has $^{55}$Fe sources mounted to a cross structure.  We used the Mn K$\alpha$ line complex from $^{55}$Fe as a diagnostic. This investigation extends the understanding of high-count-rate effects obtained from ground-based tests by analyzing distortion under realistic observational conditions.

{To facilitate reader comprehension by converting count rates into flux units, the following relation is employed:
In the closed gate valve (GV) configuration with the OPEN filter, an X-ray flux of 1 Crab ($2.4\times10^{-8}$~erg~s$^{-1}$~cm$^{-2}$ in the 2--10 keV band, assuming a photon index of 2.1; \citealt{crab}) corresponds to a count rate of 333.8 cts s$^{-1}$ for the entire Resolve array including all event grades. For one of the central four pixels, when the source is positioned at the center of the array, this corresponds to 43.2 cts s$^{-1}$ pix$^{-1}$.
Conversely, a count rate of 1 cts s$^{-1}$ pix$^{-1}$ at the central pixel is equivalent to an X-ray flux of 23.2~mCrab.
These conversions were derived using simulations performed with \texttt{heasim}. The characterization and mitigation strategies presented in this paper are valid for count rates approximately below 500 mCrab under the GV-closed configuration.}

This paper is organized as follows. Section 2 provides a description of the Resolve  instrument. Section 3 explains the observations, data screening, and data reduction. Section 4 presents the results, while Section 5 discusses their origin and implications. Finally, Section 6 concludes the paper with a summary of key findings.
Throughout this paper, errors correspond to 1$\sigma$ statistical uncertainty.

\section{Instrument}
In this section, we briefly describe the Resolve detector components and signal chain. 
It is also noted that a general description of Resolve can be found in \citet{Sato2023} and \citet{kelley2024}. The PSP is described in \citet{Ishisaki2018}, and 
the instrument characteristics critical for determining the high-count-rate effects are explained in detail in \citet{mizumoto2025}. 

The Resolve detector system consists of a 6$\times$6-pixel microcalorimeter array (PIXEL 0--35), with each pixel comprising an ion-implanted Si thermometer paired with a HgTe X-ray absorber \citep{kilbourne18b}.
Each pixel incorporates micro-machined silicon thermal links to the thicker silicon of the detector chip that is maintained at 50 mK by a multi-stage cryogenic system.
A dedicated calibration pixel located outside the aperture (PIXEL 12), illuminated by a collimated $^{55}$Fe source mounted within the detector housing, allows continuous monitoring of common-mode energy-scale variations at 6 keV.
The system includes an anti-coincidence (anti-co) detector placed behind the array to suppress background noise due to cosmic rays  \citep{kilbourne18b, porter18}. 

Signals from each pixel are processed by the Xbox analog electronics module, which amplifies, filters, and digitizes them \citep{Kelley2016}. The digital signals are then relayed to the onboard PSP \citep{Ishisaki2018} for event detection and characterization. The PSP consists of two units (PSP-A and PSP-B), each responsible for half the array, and is equipped with Field Programmable Gate Arrays (FPGAs) for pulse detection and two CPUs for advanced event processing. 
Each of the four CPU boards (PSP-A0, A1, B0, and B1) manages events from a specific quadrant, which is a sequential group of nine pixels: PSP-A0 for PIXEL 0--8, A1 for PIXEL 9--17 and one anti-co readout channel (A), B0 for PIXEL 18--26, and B1 for PIXEL 27--35 and the second anti-co readout channel (B). 
The PSP design satisfies the requirement to process $>$150 cts s$^{-1}$ from an X-ray source in combination with a background rate across the array of $>50$ cts s$^{-1}$; in other words, each CPU can handle  at least $50$~cts~s$^{-1}$~quadrant$^{-1}$ (e.g., \citealt{Ishisaki2018,mizumoto2025}).

When the event rate exceeds this limit, CPU overload, known as ``PSP overflow'', occurs. In such cases, some events are discarded, leading to event loss. Each pixel is assigned its own buffer, and event loss is more pronounced in pixels with higher event rates. The PSP records the affected pixels and the time intervals of event loss. By incorporating these event-loss interval into the analysis, the effective exposure time for each pixel can be adjusted, enabling accurate recovery of the original event rate.

Events are categorized into grades in the PSP: High-resolution Primary (Hp), Mid-resolution Primary/Secondary (Mp/Ms), and Low-resolution Primary/Secondary (Lp/Ls). This classification depends on how close the events are in time to an adjacent event of the same pixel, and 
the fractional ratios are a function of the incoming rate \citep{Ishisaki2018}. As the incoming rate increases, the fraction of the Hp events, which have the highest energy resolution, decreases. If the incoming photons are Poisson distributed, the Hp count rate has a peak of 2.2 cts s$^{-1}$ pix$^{-1}$ when the incoming rate is 6.1 cts s$^{-1}$ pix$^{-1}$. If the incoming rate exceeds this value, the Hp count rate drops. 
For very high count rate sources, the Hp spectrum will have few counts.

To reconstruct the time-dependent energy scale, Resolve employs onboard calibration sources and an energy correction algorithm \citep{porter16, porter2024,cucchetti2018,eckart18,cucchetti2024,smith2023,eckart2025}. This algorithm models time-dependent gain change as variations in an ``effective temperature,'' allowing dynamic adjustments to energy-scale calibration.
In other words, while the energy scale varies among different pixels and over time, due to various factors such as the heat sink temperature, the local radiation temperature, and the temperature of the readout electronics,
these changes are parameterized using a single variable, the effective temperature, in the algorithm.
The onboard $^{55}$Fe sources periodically irradiate the pixels, enabling energy-scale monitoring using the Mn K$\alpha$ emission lines. This approach ensures sub-eV accuracy in the energy-scale reconstruction over the science waveband, meeting the requirements of $\pm$2~eV for high-resolution spectroscopy \citep{porter2024, eckart2025}.

\section{Observations and data reduction}
\subsection{Observations}
\begin{figure}
 \begin{center}
  \includegraphics[width=\columnwidth]{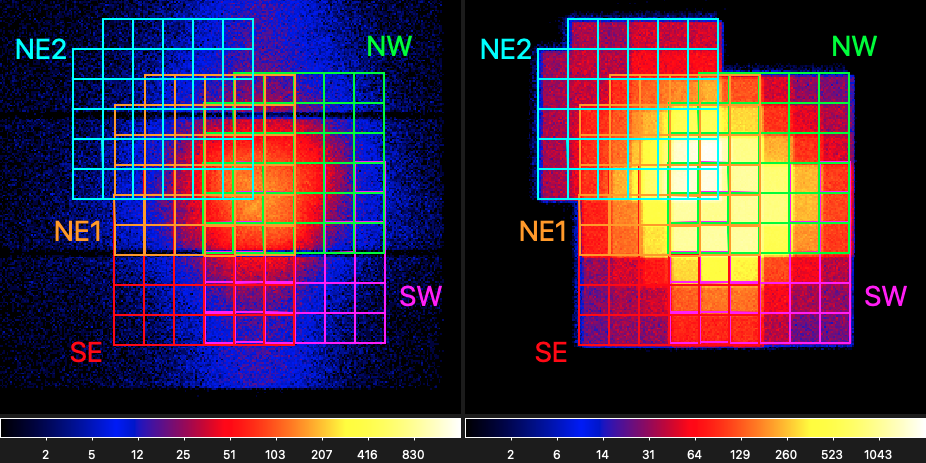} 
 \end{center}
\caption{Images of the Crab observations in sky coordinates. The left panel shows the Xtend image for 100006010 (NE1), with the Resolve field of view overplotted. Some out-of-time events are seen in the Xtend image. The right panel shows the Resolve image which is summed from all the observations. 
}\label{fig:ds9}
\end{figure}

XRISM observations of the Crab Nebula were conducted from 2024 March 18 to 20 at five offset positions (NE2, NE1, NW, SE, and SW). The observation log is summarized in Table \ref{tab:obslog} and the fields of view are shown in Figure \ref{fig:ds9}.
The GV, equipped with a $\sim$250 \si{\micro m} thick beryllium window \citep{Midooka2021}, remained closed during the observations,
which limits the band pass above $\sim$1.8~keV.
{The count rate is drastically reduced by the closed GV.}

\begin{table*}
  \tbl{Observation log}{%
  \begin{tabular}{cccccccc}
      \hline
      OBSID & Name\footnotemark[$*$] & RA\footnotemark[$\dag$] & Dec\footnotemark[$\dag$] & Roll angle\footnotemark[$\dag$] & State date&  End date & Exposure (ks)\footnotemark[$\ddag$] \\ 
      \hline
100006010 & Crab\_NEoffset\_2 (NE2) & 83.6608 & 22.0434 & 270.0 & 2024-03-18T14:03:26 & 2024-03-19T03:32:59 & 18.6\\
100006020 & Crab\_NEoffset\_1 (NE1) & 83.6479 & 22.0273 & 270.0 & 2024-03-19T04:18:52 & 2024-03-19T11:43:18 & 17.9\\
100006030 & Crab\_NWoffset (NW) & 83.6207 & 22.0280 & 270.0 & 2024-03-19T12:18:29 & 2024-03-19T18:07:09 & 13.5\\
100006040 & Crab\_SEoffset (SE)& 83.6482 & 22.0018 & 270.0 & 2024-03-19T18:42:12 & 2024-03-20T04:37:59 & 15.3\\
100006050 & Crab\_SWoffset (SW) & 83.6203 & 22.0025 & 270.0 & 2024-03-20T04:38:33 & 2024-03-20T10:06:50 & 13.1\\
      \hline
    \end{tabular}}\label{tab:obslog}
\begin{tabnote}
\footnotemark[$*$] In other parts of this paper, abbreviations in parentheses are used.\\ 
\footnotemark[$\dag$] Nominal angle (deg)  \\ 
\footnotemark[$\ddag$] Before pixel-GTI correction. \\ 
\end{tabnote}
\end{table*}

\subsection{Data screening and energy scale calibration}
Data reduction was performed with software
version ``004\_002.15Oct2023\_Build7.011'' and XRISM CALDB version 20240815. Publicly available {\tt ftools} packages were used for pipeline data reduction and data analysis.
A cleaned event file was generated by applying two types of screening: Good-Time-Interval (GTI) screening and event screening.
We have applied these standard screening criteria for the Crab data.

Throughout the observations, the FW $^{55}$Fe sources were in the aperture, irradiating the entire detector array as well as X-rays from the Crab Nebula. 
Fiducial points for the effective temperature determination were established approximately every 96 minutes during Earth occultation ($\sim$30 minutes in duration), resulting in 26 fiducial points. 
We created a spectrum during each fiducial point, collecting the data during when the elevation angle from the Earth limb is less than $-5$ degrees. Other screening criteria are the same as the standard one. 

\begin{figure*}
 \begin{center}
  \includegraphics[width=18cm]{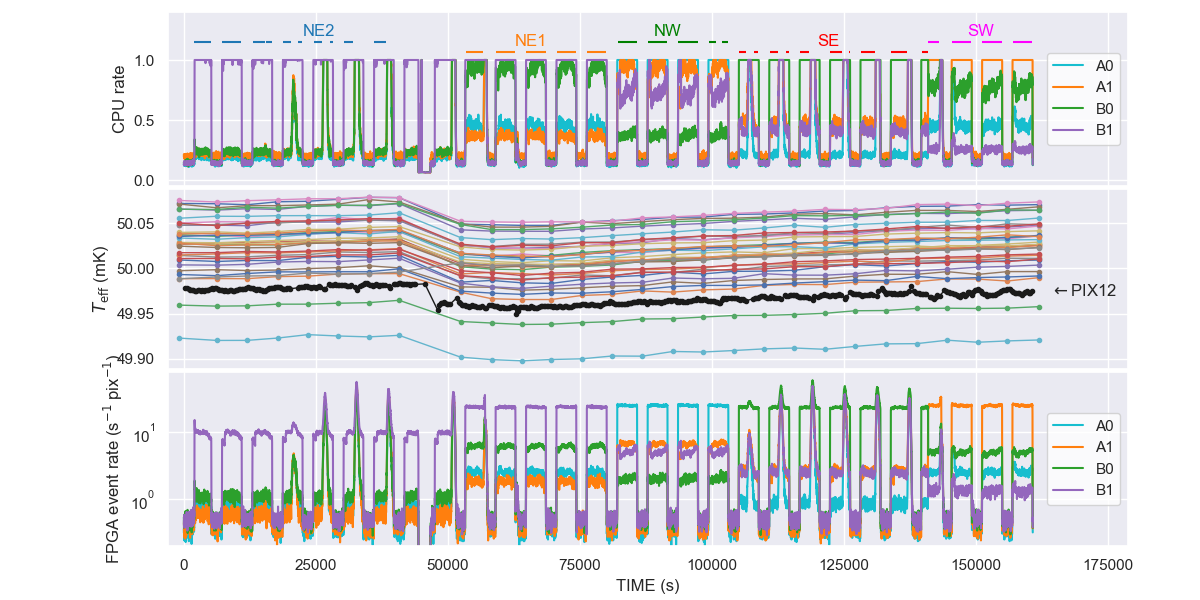} 
 \end{center}
\caption{Rate of CPU consumption, the effective temperature, and the rate of FPGA-detected event candidate with all the grades averaged in the quadrant.
Cyan, orange, green, and purple color in the data plot show PSP-A0, A1, B0, and B1, respectively. 
In the effective temperature panel, the black points are for the calibration pixel, and the colored ones are for the other pixels at the fiducial points.   
  The horizontal lines in the top panel show the GTI for the cleaned event file, before the pixel GTI correction. 

}\label{fig:psptrend}
\end{figure*}

The effective temperature for each pixel and each fiducial point 
was calculated using the {\tt rslgain} command.
Its variation is illustrated in the middle panel of Figure \ref{fig:psptrend}. We used only the Hp events for the energy-scale determination.
Photon energy is assigned to each event using the fiducial curves based on the ground calibration and the standard nonlinear energy scale interpolation method, using the {\tt rslpha2pi} command. 
We also used the fiducial points during the last Earth occultation just before the NE2 observation (OBSID=100001010) and the first one just after the SW observation (OBSID=100007010).
The last Earth occultation in NE2 (TIME$\simeq$47000~s in Figure \ref{fig:psptrend}) occurred just after the Adiabatic Demagnetization Refrigerator (ADR) recycle with a control setpoint temperature anomaly \citep{chiao2025}, so that we were not able to utilize this fiducial point. Therefore, we did not use the last 10000~s (TIME$\simeq$40000--50000~s) for NE2 in the cleaned event file.

\subsection{Exposure time correction}
The upper panel of Figure \ref{fig:psptrend} represents the average CPU consumption rate for each quadrant. As the count rate increases, the CPU consumption rate also rises. When the CPU rate exceeds unity, not all events can be processed by the PSP, leading to PSP overflow. 
The lower panel of Figure \ref{fig:psptrend} is the rate of FPGA-triggered event candidates, which serves as a proxy for the true count rate. The spike-like increases in the rate are caused by passages through the SAA, which were excluded during the GTI screening.

\begin{figure*}
 \begin{center}
  \includegraphics[width=15.5cm]{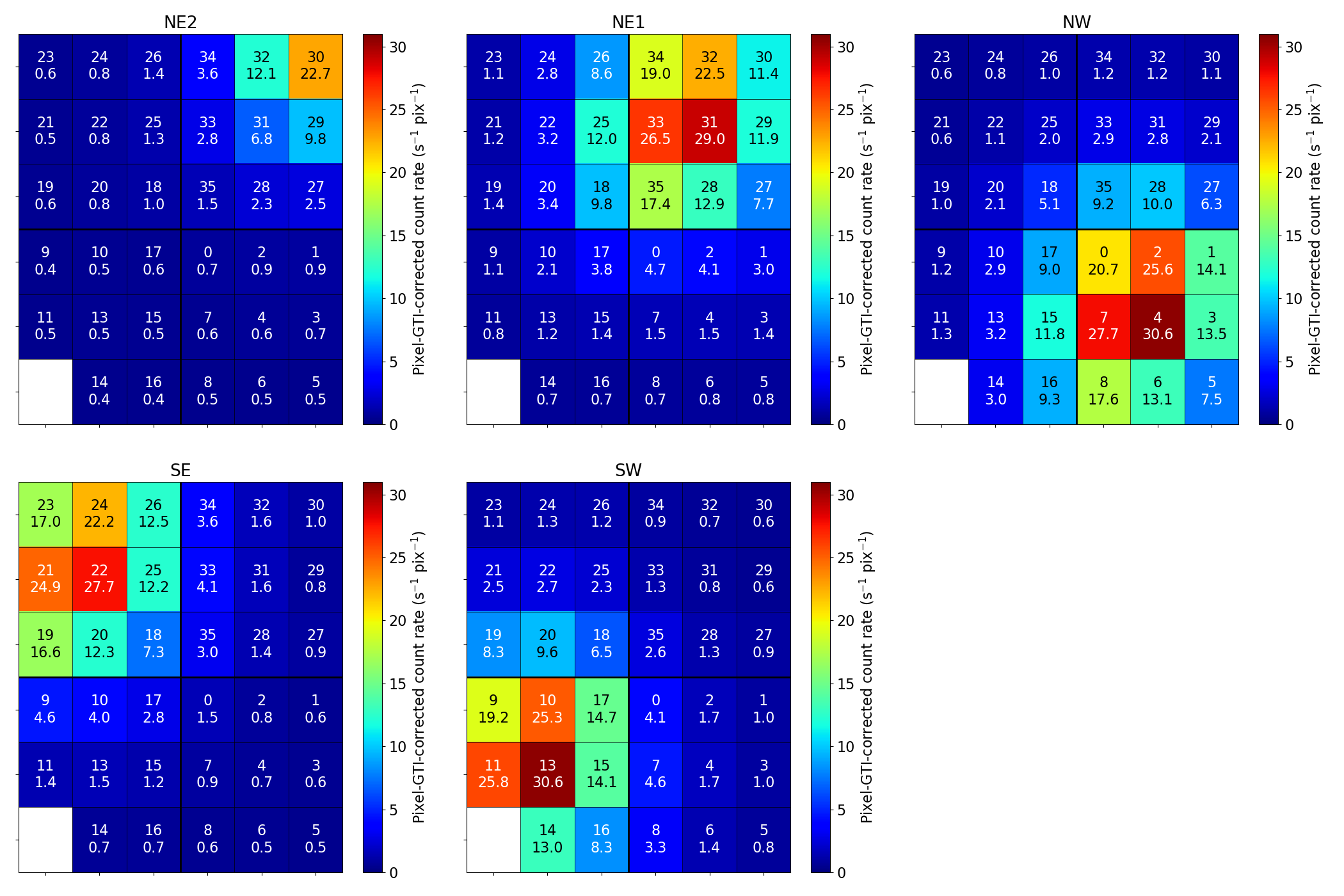} 
 \end{center}
\caption{The pixel-GTI-corrected count rate map for each observation. All event grades are represented. The upper and lower numbers are the pixel number and the count rate, respectively. 
}\label{fig:cr2_all}
\end{figure*}

When the PSP overflow occurs, the bad time intervals in which all events are discarded are set for each pixel. By subtracting it from the general GTI, the ``pixel GTI'' is made. Its length differs among pixels, so that the effective exposure time varies for each pixel. However, this kind of exposure-time correction is not performed in the pre-pipeline and pipeline processing. Therefore this correction must be applied in post-processing {(see \citealt{tsujimoto18c} for the in-orbit Hitomi/SXS results)}. 
Figure \ref{fig:cr2_all} shows the ``pixel-GTI-corrected'' count rate map for each observation, in which the exposure-time corrections due to the event loss were applied.
In the NE1 observation, the pixel-GTI-corrected count rate in the PIXEL 27--35 quadrant is 158.2 cts s$^{-1}$, exceeding the maximum processing capacity in the PSP (50 cts s$^{-1}$). Therefore PSP overflow has occurred in this quadrant and some events have been discarded. See Appendix \ref{a1} for a detailed explanation of calculating the pixel-GTI-corrected count rate.
It should be noted that other mechanisms may remain to distort the count rate, such as dead time due to pile-up effects \citep{Ishisaki2018,mizumoto2025} and false detection of secondary pulses (see Section 6.3 in the XRISM ABC Guide v1.0\footnote{\url{https://heasarc.gsfc.nasa.gov/docs/xrism/analysis/abc_guide/xrism_abc.pdf}}), which is beyond the scope of this paper.

\subsection{Cross-talk cut}
Electrical cross-talk arises from interactions between pixels with adjacent read-out wires, producing pseudo-pulses with small pulse heights, here referred to as ``cross-talk children.'' 
When X-ray events occur close in time on neighboring channels, the mutual cross talk between them, depending on the energies, temporal displacement, and magnitude of the coupling, can alter the inferred energy of each event.
Depending on the parent event’s energy and the temporal displacement between events, this contamination can alter the inferred X-ray energy. The degradation in energy resolution becomes more pronounced at higher count rates due to the increased likelihood of overlapping events.
In this study, events contaminated by untriggered electrical cross-talk children were identified when a pulse in PIXEL $i$ occured within $\pm 25$~ms of another pulse in PIXEL $i\pm1$ within the same quadrant. Events meeting these criteria were excluded during a process termed the ``cross-talk cut'' \citep{mizumoto2025}.

The cross-talk cut described above creates a GTI filter in pixel $i$ based on the arrival times of events in pixels $i\pm1$ in the same quadrant. During PSP overflow, some events in pixel $i\pm1$ are discarded and creating a GTI to filter out their cross-talk children in pixel $i$ is not possible. To address this, periods of event loss in {pixels $i-1$ or $i+1$} are excluded when creating the GTI for pixel $i$.
{Note that this cut, which removes pairs of potentially contaminated X-ray events, is labeled ``long'' cross-talk screening (see also \citealt{mochizuki2025}).   ``Short'' cross-talk screening, which removes triggered cross-talk children while retaining their parents, is needed only for reducing the background below 0.3 keV and is not addressed in this paper.}

\subsection{How to evaluate energy scale and resolution}
To study the variation in energy scale and resolution, we created a Hp spectrum in the interval 5.85~keV to 5.93~keV (around the Mn K$\alpha$ lines) for each pixel and observation. 
{
The model included eight Lorentzian components, following \citet{holzer1997} with corrections to known errors\footnote{\url{https://heasarc.gsfc.nasa.gov/docs/hitomi/calib/caldb_doc/asth_sxs_caldb_linefit_v20161223.pdf}}, and incorporates Gaussian broadening to model the Mn K$\alpha^{1,2}$ lines}, along with a power-law component with a fixed photon index of 2 to describe the Crab spectrum. The model fitting was performed in XSPEC with C statistics \citep{cash1979}, using a diagonal response file. The ``gain fit'' command was employed to determine the energy offset at 6 keV. The model contains four free parameters: Gaussian broadening, normalization of the Lorentzians, normalization of the power law, and the energy offset at 6 keV.  These parameters were optimized for each pixel to analyze energy-scale shifts and energy-resolution changes across the observations at 6 keV.
We did not use PIXEL 27 for the spectral analysis, which is subject to random gain jumps of unknown origin \citep{porter2024}.

\section{Results}
\subsection{Energy-scale shift at 6~keV as a function of count rate}
\begin{figure}
 \begin{center}
  \includegraphics[width=0.99\linewidth]{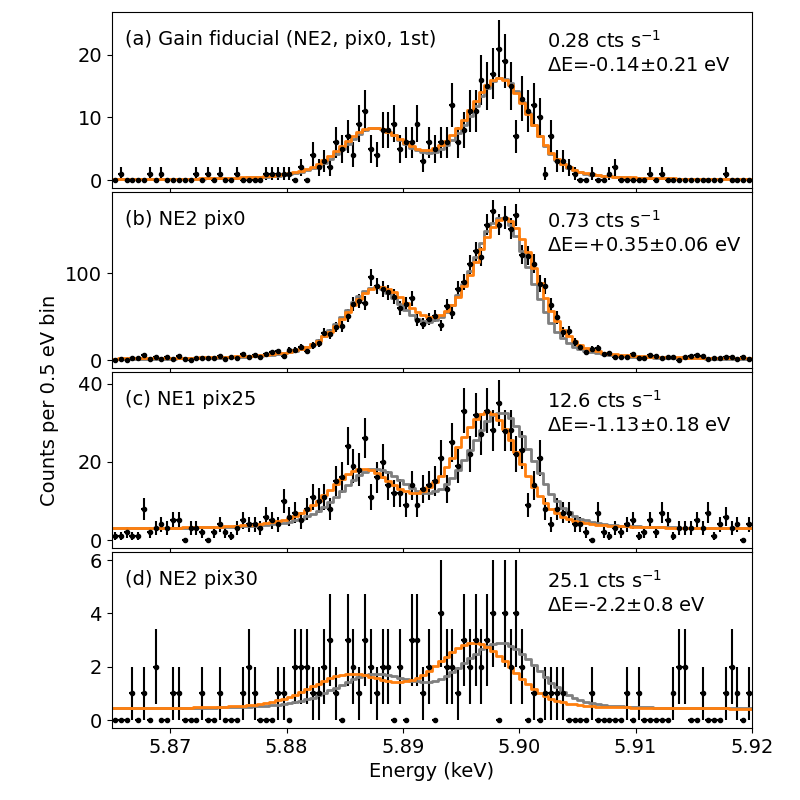} 
 \end{center}
\caption{Examples for the Mn K$\alpha$ model fitting. The black bins show the spectral data. The orange line is the best fit model. The gray line is also the best fit but the energy offset is frozen at 0. (a) is the spectrum for one of the fiducial points. (b)--(d) are the ones for the Crab observation, whose count rate is small, high, and very high, respectively.

}\label{fig:spec_comp}
\end{figure}

The energy-scale shift and energy resolution at 6 keV were measured for each pixel and observation. Figure \ref{fig:spec_comp} shows representative examples of the model fit under varying count rates.
Panel (a) shows the spectrum measured at a fiducial point during which the view to the Crab was occulted.  As this point was used to correct the energy scale, it does not exhibit a significant energy offset.
Panels (b)--(d) show the Mn K$\alpha$ lines with increasing count rates from the Crab observations. At low count rates (panel b), a slight positive energy-scale shift at 6 keV is observed. As the count rate increases (panels c and d), the energy-scale shift at  6 keV becomes increasingly negative.

\begin{figure}
 \begin{center}
  \includegraphics[width=0.95\linewidth]{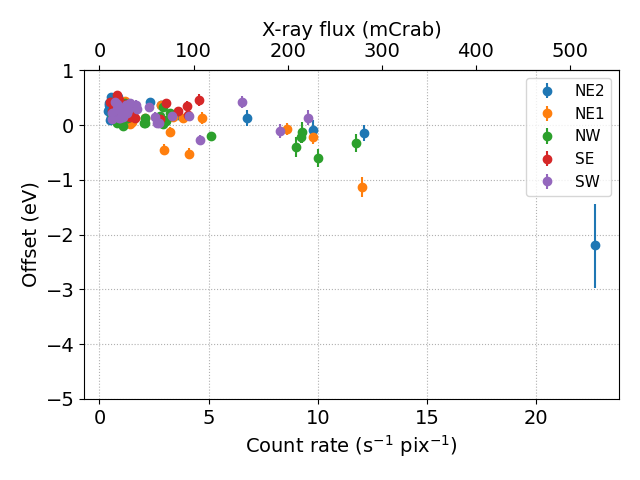} 
  \includegraphics[width=0.95\linewidth]{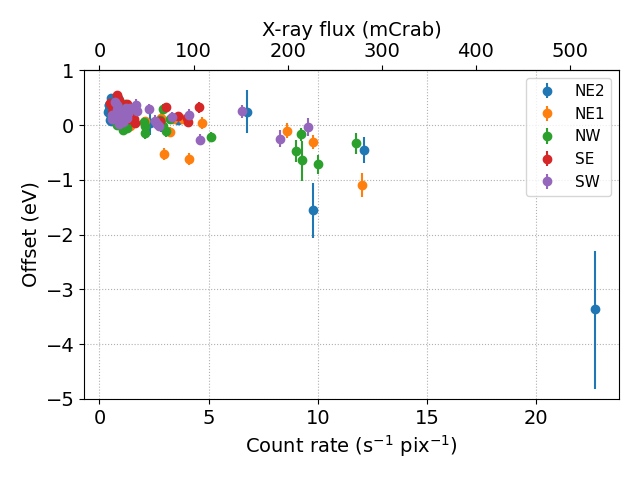} 
  \includegraphics[width=0.95\linewidth]{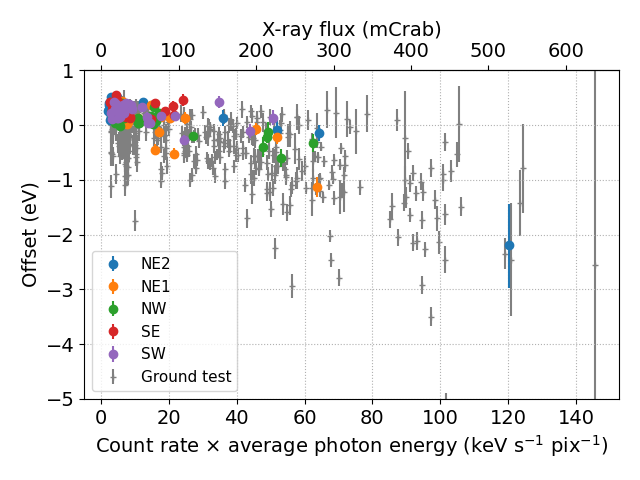} 
 \end{center}
\caption{(Upper) Energy offset at 6 keV versus the pixel-GTI-corrected count rate for the Crab nebula observation. Different color shows different observation. 
(Middle) Same as the upper panel,  but after the cross-talk cut.
(Lower) Same as the upper panel, but the horizontal axis is pixel-GTI-corrected count rate times the photon energy.  The ground test results are also shown in gray \citep{mizumoto2025}. As the horizontal axis increases, the thermal input from the incoming photons gets larger.
In all the panels, the upper horizontal axis indicates the corresponding X-ray flux in mCrab.
}\label{fig:cr_offset}
\end{figure}

The upper panel of Figure \ref{fig:cr_offset} presents the relationship between the energy offset at 6 keV and the pixel-GTI-corrected count rate. Each data point represents a pixel. Pixels with extremely low number of Hp counts due to the PSP overflow and small Hp fraction (namely, PIXEL28--35 in NE1, 0--8 in NW, 18--26 in SE, and 9--17 in SW) were not used. 
The energy offset at 6~keV is slightly positive at low count rates ($\lesssim6$~cts~s$^{-1}$~pix$^{-1}$), and becomes increasingly negative as the count rate rises. 
This trend is maintained even after the cross-talk cut was performed {(the middle panel of Figure \ref{fig:cr_offset}).
The difference between the relationship before and after the cross-talk cut is negligible, at most 0.05 eV, since pulse contamination due to the cross talk children can increase or decrease the original X-ray energy, depending on the energy of the parent pulse and the displacement between pulses \citep{mizumoto2025}. }

\citet{mizumoto2025} studied the energy offset under the high count rate situation in the ground-based test, and reported negative energy offsets at 6 keV in the count rate regime.
The lower panel of Figure \ref{fig:cr_offset} shows the energy offset at 6 keV as a function of the total energy of incoming photons, for both of the ground-based test (gray) and in-flight observation (color). 
In the ground-based test, the whole detector array was illuminated uniformly. In this case, the detector frame between and around the pixels in the detector array can be locally heated, and thus the heat sink temperature in the array can slightly increase. This can {cause} a negative energy offset.
On the other hand, in the in-flight observation, the X-ray illumination is not uniform across the array, resulting in a complex temperature distribution across the local heat sink of each pixel.
While we cannot directly compare the energy offset value between the ground-based test and the in-flight observation since the immediate thermal environment of each pixel is different and the dispersion in the ground-based test data is very large,
they share the same trend of more negative offset at higher input power.
At least, it is clear that the energy scale can vary across the array in observations of very bright point sources.

The positive offset at low count rates is likely caused by orbital variation in the Resolve electronics and sparse fiducial points.
This mechanism is discussed further in Sections 5.1.

\subsection{Energy resolution degradation at 6 keV and recovery}
\begin{figure}
 \begin{center}
 \includegraphics[width=0.95\linewidth]{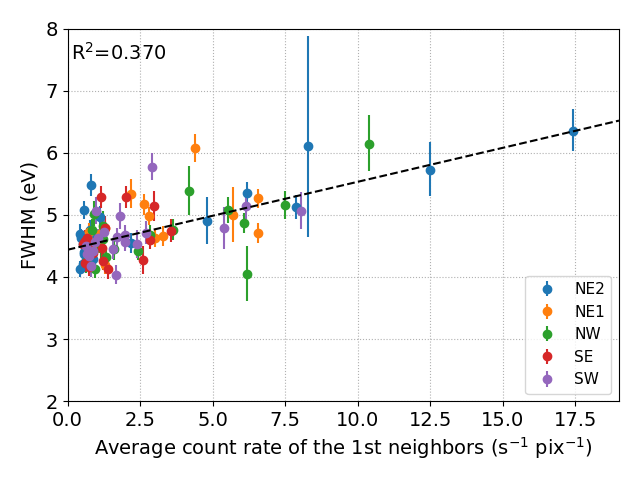} 
  \includegraphics[width=0.95\linewidth]{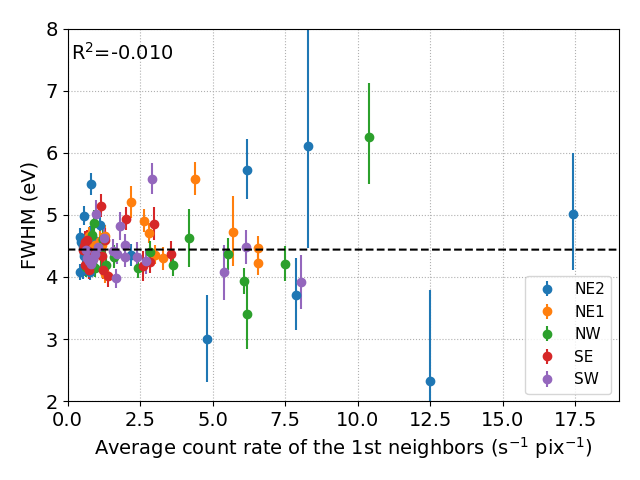} 
 \end{center}
\caption{Energy resolution (FWHM) at 6 keV versus the pixel-GTI-corrected count rate. The upper panel is the result using the data before the cross-talk cut, and the lower, after the cut. The black dashed line shows the best fit model (Eq.~\ref{eq2}).
}\label{fig:cr_fwhm}
\end{figure}

The energy resolution, represented as the Full-Width at Half-Maximum (FWHM) of the Gaussian instrumental function, was evaluated across varying pixel-GTI-corrected count rates. As shown in the upper panel of Figure \ref{fig:cr_fwhm},  the energy resolution at 6 keV degrades with increasing count rates in neighboring pixels due to electrical cross talk. This relationship was modeled with a linear function:
\begin{equation}
  y=ax+b    \label{eq2},
\end{equation}
where $y$ is FWHM (eV) at 6 keV and $x$ is the Pixel-GTI-corrected count rate (s$^{-1}$ pix$^{-1}$).
The fit yields $a=0.109\pm0.014$ and $b=4.44\pm0.04$. The adjusted coefficient of determination ($R^2$) is 0.370. 
After applying the cross-talk cut, however, the correlation between neighboring pixel count rates and the energy resolution at 6 keV becomes negligible, as shown in the lower panel of Figure \ref{fig:cr_fwhm}. The best-fit parameters are $a=0.00\pm0.02$ and $b=4.44\pm0.05$, with $R^2=-0.010$. This result demonstrates that the cross-talk cut effectively restores the energy resolution to levels consistent with ground-based testing \citep{mizumoto2025}.

\begin{figure}
 \begin{center}
 \includegraphics[width=0.95\linewidth]{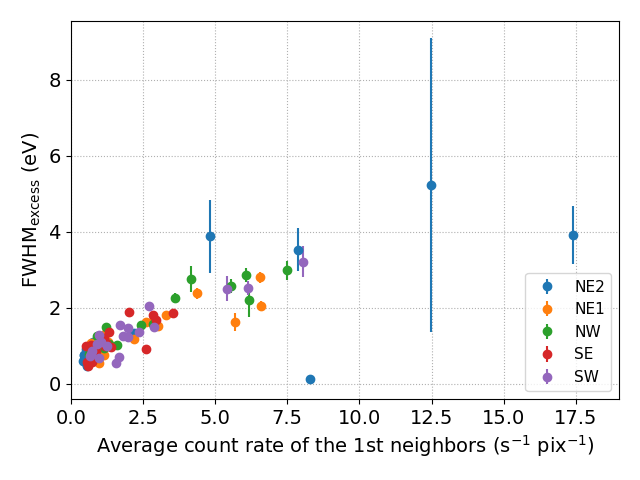} 
 \end{center}
\caption{
The excess of energy resolution (FWHM) versus the pixel-GTI-corrected count rate. The data in which $\mathrm{FWHM_{withXtalk}}>\mathrm{FWHM_{noXtalk}}$ are only shown.}
\label{fig:cr_fwhm_excess}
\end{figure}

{
The excess broadening due to the cross-talk effect (excess FWHM) is defined to be  
\begin{equation}
 \mathrm{FWHM_{excess}}=(\mathrm{FWHM_{withXtalk}}^2-\mathrm{FWHM_{noXtalk}}^2)^{1/2},
\end{equation}
where $\mathrm{FWHM_{withXtalk}}$ is FWHM before the cross-talk cut and $\mathrm{FWHM_{noXtalk}}$ is the one after the cross-talk cut, as shown in Equation (2) in \citealt{mizumoto2025}.
Figure \ref{fig:cr_fwhm_excess} shows the excess FWHM derived from our observations. Note that we fitted a straight line to FWHM vs count rate instead of excess FWHM in order to apply the same fit after applying the cross-talk cut, for which the calculated excess is often imaginary.}

\begin{figure}
 \begin{center}
 \includegraphics[width=0.95\linewidth]{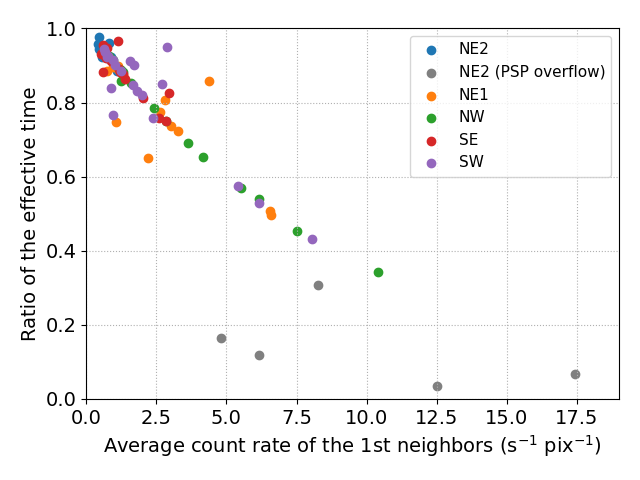} 
 \end{center}
\caption
{The ratio of the effective exposure time after/before the cross talk cut.}

\label{fig:cr_xttime}
\end{figure}

{
When the cross-talk cut is applied, some exposure time is lost in exchange for restored energy resolution.
Figure \ref{fig:cr_xttime} shows the result, where the ratio of the effective exposure time after and before the cross-talk cut is plotted against the count rate for each pixel.
In some pixels in NE2, where the PSP overflow occurs,
periods of event loss in {pixels $i-1$ or $i+1$} are excluded when creating the GTI for pixel $i$ (see Section 3.4). Therefore, the remaining time after event loss is shorter than the other pixels with similar count rates. In any case, the cross-talk cut loses more effective exposure time with the larger count rates in the neighboring pixels.
}

\section{Discussion}

\subsection{Variation of the energy offset during the orbital period of the satellite}
\begin{figure}
 \begin{center}
  \includegraphics[width=0.95\linewidth]{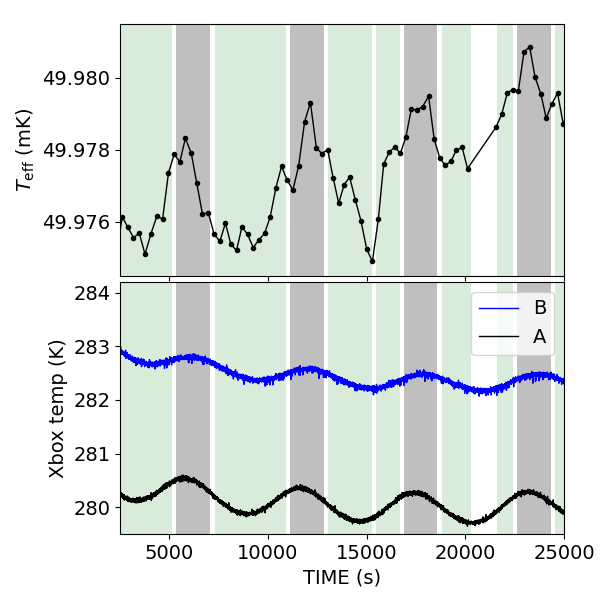} 
 \end{center}
\caption{The change of the effective temperature in the calibration pixel (upper) and the temperature on the two Xbox amplifier boards. The light-green-shaded area shows GTI of the cleaned event file, while the dark-gray-shaded area shows the one for the fiducial data.
}\label{fig:psplimit_enlarge}
\end{figure}

At low count rates, a positive energy offset of approximately $+$0.2 eV was observed at 6 keV. This offset is primarily attributed to to the complex interplay of various factors, including variations in the gain of the Xbox amplifier and sparse sampling of the orbital energy-scale variation during fiducial points.
Figure \ref{fig:psplimit_enlarge} illustrates the change of effective temperature in the calibration pixel and temperature on the two Xbox amplifier boards. The effective temperature variation was calculated for the entire observation period, revealing a periodic fluctuation with a timescale matching the orbital period of the satellite ($\sim$96 minutes), with the same periodicity as the lower panel.
This behavior is consistent with Hitomi/Soft X-ray Spectrometer observations (e.g., Figure 1 in \citealt{sawada2022}).
It should be noted that the temperature of the amplifier boards is dominated by various factors, such as the detector bias supply and the temperature of the ADR controller. In any case, periodic changes associated with the satellite's orbital motion appear as changes in the effective temperature.

During our observations, the effective temperature of the calibration peaked near the time of Earth occultation. 
Energy-scale correction was performed by interpolating the effective temperature at the fiducial points, resulting in cleaned event data showing temperatures up to 3 \si{\micro K} lower than the fiducial point, with an average difference of approximately 1 \si{\micro K}. Based on the effective temperature vs energy plots \citep{eckart2025,porter2024}, a temperature drop of 1 \si{\micro K} corresponds to a positive energy-scale shift of $+$0.2 eV at 6 keV, which aligns perfectly with the observed values.

When the Earth occultation coincides with the energy-scale oscillation peak, the energy-scale shift is almost at its maximum. Even in this case, the shift is sufficiently small. While similar energy-scale shifts could theoretically occur in any observations, it is important to note that the energy-scale oscillation peak does not always align with the Earth occultation. 
In fact, \citet{porter2024} shows that the energy scale during the fiducial points and the one except for the fiducial points are very consistent, within 0.2~eV error and many are less than 0.1~eV at 6 keV, for the $\sim80$ celestrial targets, and that even across the 2--10 keV band, the energy scale error is less than 1 eV.

\subsection{Application for point sources}
The majority of potential high count rate sources for XRISM observations will be Galactic compact objects. Here, we discuss how the energy-scale shift and energy resolution degradation affect (or can be neglected in) the analysis for these sources.

GX 13+1, a low-mass X-ray binary with a neutron star and a solar-mass companion \citep{iaria2014}, serves as a test case. This source, with an X-ray flux near half the Eddington luminosity (e.g., \citealt{Tomaru2020b}), exhibits deep absorption lines from accretion disc winds (e.g., \citealt{Ueda2004}). XRISM observed this source on 2024 February 25, as part of the performance verification observations (OBSID=300036010,  \citealt{xrism_gx13}). 

\begin{figure}
 \begin{center}
  \includegraphics[width=0.95\linewidth]{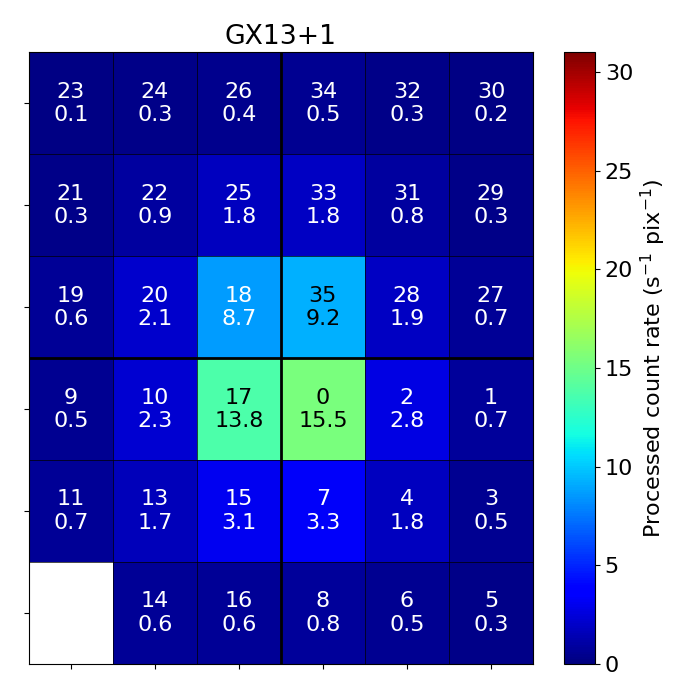} 
 \end{center}
\caption{The processed count rate for the GX 13+1 observation.

}\label{fig:cr_gx13}
\end{figure}

Figure \ref{fig:cr_gx13} shows the count rate map. The count rate per quadrant is 26.2, 23.3, 15.3, and 15.8 cts~s$^{-1}$ for PSP-A0, A1, B0, and B1, respectively. These rates are below the maximum throughput of 50 cts~s$^{-1}$~quadrant$^{-1}$, ensuring no PSP overflow.
Spectra were extracted for the inner (brighter) 4 pixels (PIXEL0, 17, 18, 35) and the outer (fainter) pixels, excluding the inner 4 pixels in addition to pixels 12 and 27.
The screening criteria, energy correction methods, and response files used are the same as in \citet{xrism_gx13}, except that the response files were individually generated for each spectrum.
The average photon energy for the GX 13+1 observation is $E_\mathrm{ave}=5.3$~keV, which is similar to the one for the Crab, $E_\mathrm{ave}=5.0$~keV.

\begin{figure}
 \begin{center}
  \includegraphics[width=0.95\linewidth]{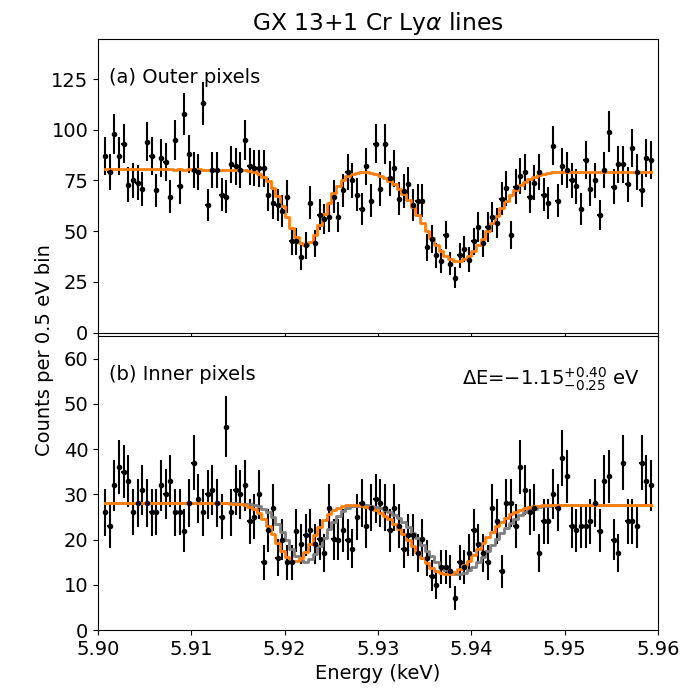} 
 \end{center}
\caption{Comparison of the Hp spectra for GX 13+1 between the outer pixels and the inner pixels. The orange line is the best fit, while the gray line is also the one but the energy offset is not allowed. 

}\label{fig:gx13_speccomp}
\end{figure}

\subsubsection{{Energy-scale shift in GX 13+1}}
{First, we investigate the energy-scale shift in each pixel.}
The inner pixels have count rates of 8.7--15.5 cts~s$^{-1}$~pix$^{-1}$.
As seen in section 4.1, some negative energy offset with an order of sub-eV or eV can be expected in the brighter pixels at 6~keV.
Figure \ref{fig:gx13_speccomp} shows the GX 13+1 spectrum around the Cr Ly$\alpha^{1,2}$ lines, which are strong but not saturated, and close in energy to the Mn K$\alpha$ calibration lines ($\sim6$~keV).
The upper panel shows the spectrum for the outer pixels, while the lower panel, for the inner pixels. 
The model fitting was performed within 5.90--5.96 keV, with the model of power law and two negative Gaussians. The difference between the energy centroids of the two absorption lines and the width of each absorption line were tied.
The energy offset with $-1.15_{-0.25}^{+0.40}$~eV is detected at 6 keV in the inner pixels, compared to the outer ones. 
This energy offset could be erroneously interpreted as a velocity difference of $\sim$+50~km~s$^{-1}$ if flux-dependent effects were not considered.

This shift is sufficiently low and users can analyze this data without any additional treatment for most of the science cases.
However, it should be noted that,
if the precise determination of wind velocity within the accuracy of the order of tens km~s$^{-1}$ has significant physical implications, it would be essential to identify potential rate-dependent energy shifts.
{Moreover, if such energy shifts across the array are not taken into account, combining spectra from all pixels may artificially broaden the spectral lines. This could affect the interpretation of line widths.}
Specifically, this would involve conducting spectral fitting for each pixel and measuring the energy offset to the central pixels.

\subsubsection{{Energy resolution degradation in GX 13+1}}
Next, we investigate the energy resolution degradation {in each pixel}.
In contrast with the Crab observation, when the point source is observed with the source position at the center of the array, the effect of the cross-talk children is small.
The most illuminated pixels are PIXEL 0, 17, 18, and 35. Their electrical neighbor pixels are PIXEL 1, 16, 19, and 34, respectively, whose count rate is as small as $\sim0.5$~cts~s$^{-1}$~pix$^{-1}$.
They have little effect on the energy resolution of the bright pixels, and the cross-talk parents in the bright pixels only affect the energy resolution in the low count rate pixels.
As a result, the energy resolution degradation is expected to be negligible.

\begin{figure}
 \begin{center}
  \includegraphics[width=0.95\linewidth]{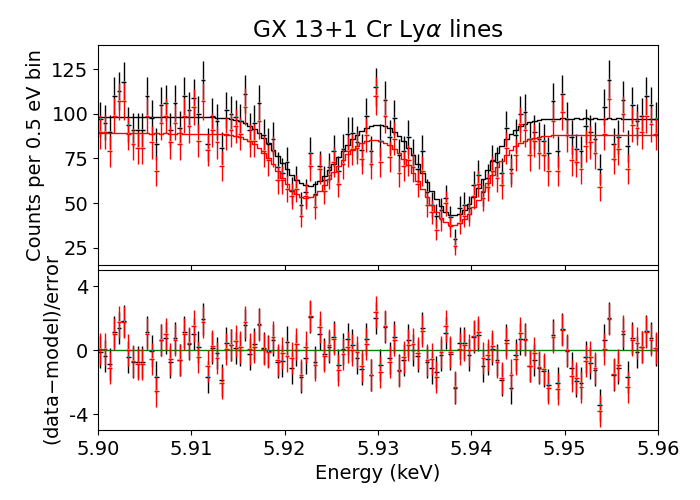}
 \end{center}
\caption{Comparison of the Hp spectra of GX 13+1 before (black) and after (red) the cross-talk cut. The lower panel shows the residuals.

}\label{fig:gx13_xtcut}
\end{figure}

We created the GX 13+1 Hp spectra and associated response files using all pixels except PIXEL 12 and 27, before and after the cross-talk cut. 
Figure \ref{fig:gx13_xtcut} shows the comparison. The fitting model is the same as in Figure \ref{fig:gx13_speccomp}, but the broadening of the Gaussians is set to be free. The derived {line widths ($\sigma$) correspond to} $129\pm6$~km~s$^{-1}$ (Cr Ly$\alpha^1$) and  $136_{-5}^{+7}$~km~s$^{-1}$ (Cr Ly$\alpha^2$) for the data before the cross-talk cut, and 
$123_{-10}^{+7}$~km~s$^{-1}$ (Cr Ly$\alpha^1$)  and $111_{-8}^{+10}$~km~s$^{-1}$ (Cr Ly$\alpha^2$) for the data after the cross-talk cut.
The turbulent velocity before and after the cross-talk cut is consistent within the $1\sigma$ error, so we cannot detect any energy resolution degradation, as expected. 

However, there are scientific cases in which a cross‐talk cut becomes necessary; for instance, when the accuracy of the centroid shift (after correcting for gain shift) is critical. 
The centroid error scales roughly as $\Delta E^{1.5}$, where $\Delta E$ is the energy resolution (FWHM), because it can be formulated by
\begin{equation}
\text{error(centroid)} \sim \frac{c \, \Delta E}{2.35 \, E \, (S/N)},
\end{equation}
where $c$ is a correction factor and $S/N$ is the signal-to-noise ratio for the line.
$S/N$ depends on $1/\sqrt{\Delta E}$ for a line weak relative to the continuum, since 
\begin{equation}
N = \sqrt{F_\mathrm{continuum} + F_\mathrm{line} + F_\mathrm{background}}, 
\end{equation}
where $F$ is the flux, and $F_\mathrm{continuum}$ and $F_\mathrm{background}$ are linear in $\Delta E$. In many cases, such as for the Fe~K lines observed on top of a strong continuum in bright sources, even small errors in the centroid can significantly affect the measurement of redshifts. Therefore, applying a cross-talk cut remains essential to ensure the accuracy of these centroid measurements.

\section{Conclusion}
We have observed the Crab Nebula using the XRISM/Resolve instrument under various offset positions and analyzed the effects of high count rates on the energy scale and energy resolution at 6 keV. Using Mn K$\alpha$ lines from the $^{55}$Fe sources, we investigated spectral distortions and demonstrated methods to mitigate these effects.
{It should be noted that the results are valid for count rates below $\sim500$~mCrab (GV closed), where sufficient statistics can be ensured.}

Our study revealed a relationship between the total energy of incoming photons and energy scale shifts at 6 keV. At higher count rates, a negative energy-scale shift was observed, likely caused by thermal effects as a result of the influx of X-ray photons. 
In contrast, at low count rates, a positive energy-scale shift of approximately $+$0.2 eV was detected at 6 keV, which can be attributed to orbital variation in Resolve electronics, which is sparsely sampled by our fiducial intervals. It should be noted that this kind of shift is usually less than this value for normal observations.

Energy resolution degradation caused by untriggered electrical cross-talk was also observed. This degradation became more pronounced with higher count rates in neighboring pixels. However, applying a cross-talk cut effectively restored the energy resolution to levels consistent with ground-based testing. These results emphasize the importance of corrective measures, when appropriate, during data analysis to ensure accurate spectral measurements.

We applied these findings to a point source, using GX 13+1 as a test case. The energy-scale shift of the inner pixels relative to the outer ones was measured as approximately 
$-$1~eV at 6 keV, corresponding to a velocity shift of $+$50~km~s$^{-1}$. This value is well within the calibration requirement. However, for precise velocity measurements with an accuracy of $\pm$10~km~s$^{-1}$, detailed pixel-specific spectral fitting and energy-offset corrections may be necessary. The energy resolution degradation for this dataset was found to be negligible, as expected, rendering cross-talk cuts unnecessary.

\begin{ack}
This work is made possible only with the contributions of the members of the XRISM \textit{Resolve} team, in-flight calibration team, and operation team.
This work was supported by 
JSPS KAKENHI grant Nos.\ JP21K13958 (M.M.), JP24K17105 (Y.K.),
Yamada Science Foundation (M.M.), and
NASA under award number 80GSFC21M0002 (K.P.).
Part of this work was performed under the auspices of the U.S. Department of Energy by Lawrence Livermore National Laboratory under Contract DE-AC52-07NA27344 (M.E.E.).
\end{ack}


\section*{Data availability} 
The XRISM in-flight data underlying this article are available in heasarc data repository (NASA/GSFC) or DARTS (JAXA/ISAS). All the data reduction tools are available as a ftools package.

\bibliography{main}{}
\bibliographystyle{aasjournal}

\appendix 

\section{Count rate correction} \label{a1}

\begin{figure*}
 \begin{center}
  \includegraphics[width=15.5cm]{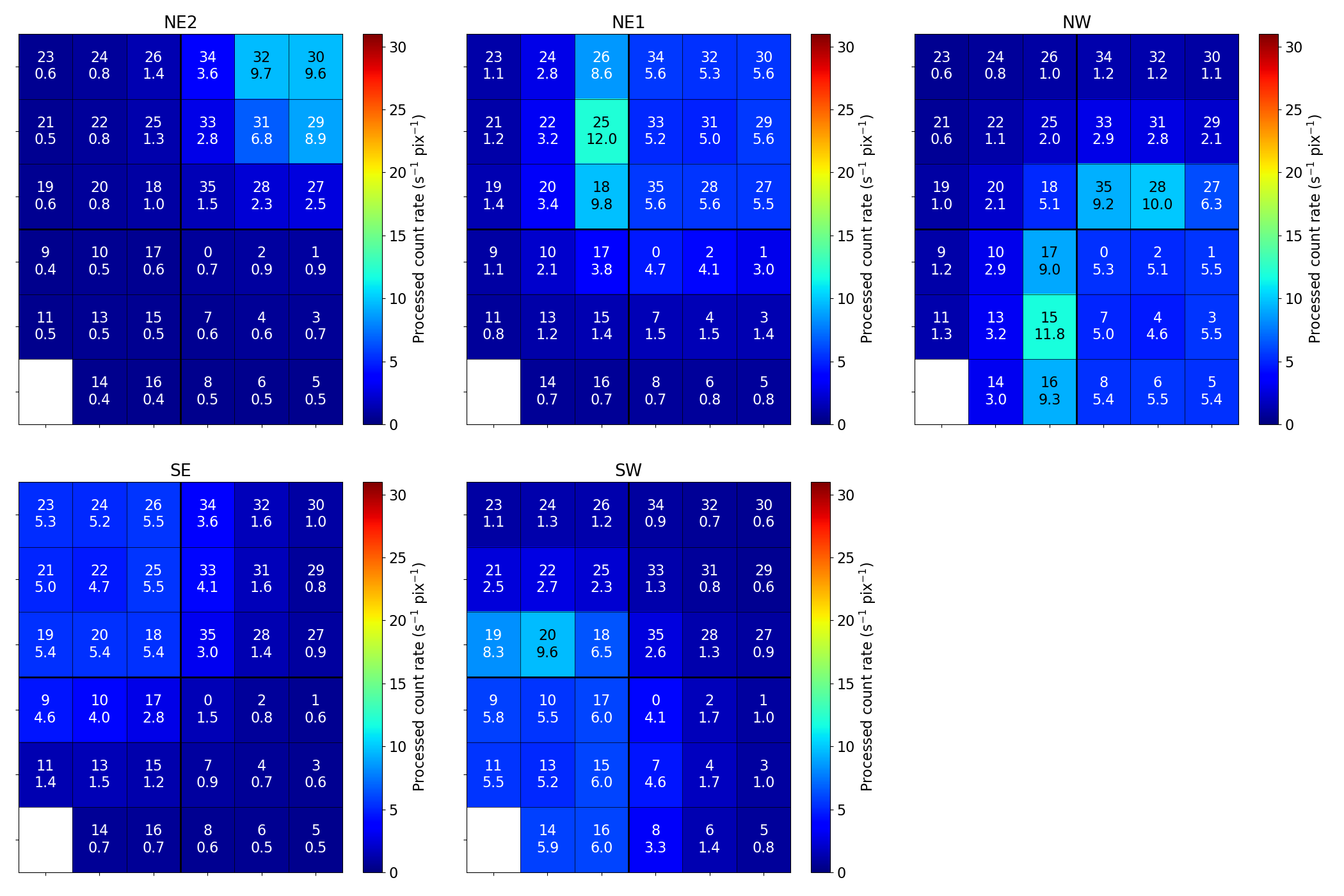} 
 \end{center}
\caption{Same as Figure \ref{fig:cr2_all}, but for the processed count rate map.

}\label{fig:cr_all}
\end{figure*}

\begin{figure*}
 \begin{center}
  \includegraphics[width=15.5cm]{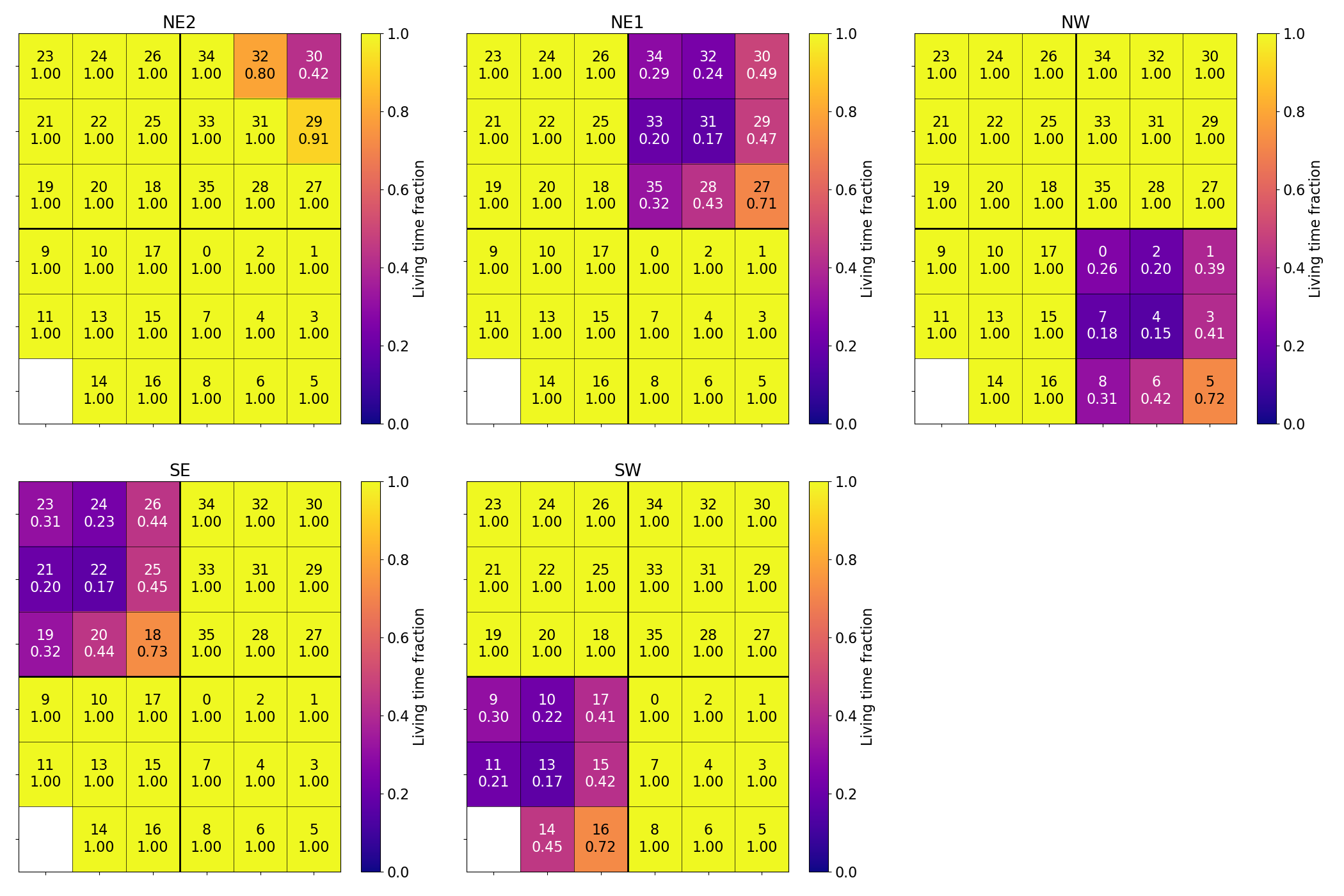} 
 \end{center}
\caption{Live time fraction map. A number less than unity indicates that a PSP overflow has occured in the pixel and thus not all the events are processed.
}\label{fig:timemap_all}
\end{figure*}

In this section, we explain how the count rate is distorted during times of PSP overflow. Figure \ref{fig:cr_all} shows the count rate for each pixel derived from the cleaned event files. For example, in the NE1 observation, PSP-B1 (PIXEL 27 to 35) experienced PSP overflow, resulting in the highest apparent count rate in adjacent pixels, such as PIXEL 25.
Figure \ref{fig:timemap_all} displays the live time fraction, which represents the fraction of GTI remaining after excluding the pixel GTIs. For the NE1 observation, PIXEL 27 to 35 exhibit live time fractions below unity, as expected. 
The PSP-processed count rate can be corrected by dividing it by the live time fraction, yielding the ``pixel-GTI-corrected'' count rate, as shown in Figure \ref{fig:cr2_all}. 
In the NE1 observation, this count rate peaks at PIXEL 31, consistent with the detector's configuration in Figure \ref{fig:ds9}.

\end{document}